\documentclass[letterpaper]{article}
\usepackage{spconf,amsmath,graphicx}

\title{Knowledge-Aided Normalized Iterative Hard Thresholding Algorithms and Applications to Sparse Reconstruction}
\name{Qianru Jiang$^{*\dagger}$, Rodrigo C.~de Lamare$^{\dagger\ddagger}$, Yuriy Zakharov$^\dagger$, Sheng Li$^{*}$ and Xiongxiong He$^{*}$ \thanks{This work was partially supported by National Science Foundation of P.R. China (Grant: 61473262, and 61503339), Zhejiang National Science Foundation (Grant: LY18F010023). The work of Q. Jiang was also supported by China Scholarship Council Funding. }}

\address{* \normalsize{\emph{College of Information Engineering, Zhejiang University of Technology, Zhejiang, People's Republic of China}}\\
    $\dagger$ \normalsize{\emph{Department of Electronic Engineering, University of York, York YO10 5DD, U.K}}\\
    $\ddagger$ \normalsize{\emph{CETUC, PUC-Rio, Rio de Janeiro 22451-900, Brazil}}\\
    \normalsize{jqr08141989@163.com, rcdl500@ohm.york.ac.uk, yury.zakharov@york.ac.uk, shengli@zjut.edu.cn, hxx@zjut.edu.cn}}
\begin{document}
%
\maketitle
\begin{abstract}
This paper deals with the problem of sparse recovery often found in
compressive sensing applications exploiting \textit{a priori}
knowledge. In particular, we present a knowledge-aided normalized
iterative hard thresholding (KA-NIHT) algorithm that exploits
information about the probabilities of nonzero entries. We also
develop a strategy to update the probabilities using a recursive
KA-NIHT (RKA-NIHT) algorithm, which results in improved recovery.
Simulation results illustrate and compare the performance of the
proposed and existing algorithms.
\end{abstract}
\section{Introduction}
\label{sec:intro}

Sparse signal recovery problems are among the fundamental problems
addressed in compressed sensing (CS) \cite{donoho2006compressed,
candes2008introduction,intadap,jio,jidf,smtvb,sjidf,l1stap} which
have been successfully utilized in many areas such as image
processing \cite{li2015designing}, wireless sensor networks
\cite{xu2015distributed}, and face recognition
\cite{wright2009robust}. Most often this problem refers to
recovering a sparse signal $\mathbf{x}$ from its linear
measurements:
\begin{equation}\label{cs}
\mathbf{y=Ax+e},
\end{equation}
where $\|\mathbf{x}\|_{0}\leq K$ (here $\|\cdot\|_{0}$ counts the number of non-zero elements), $\mathbf{A}\in\Re^{M\times N}$ with $M\leq N$, and $\mathbf{e}{\in\Re^{M\times 1}}$ is the noise which is modeled by independent and identically distributed Gaussian random variables with zero mean and variance $\sigma^{2}$.

Although most of the signals are not directly sparse, they can be
sparsely represented by a well-chosen dictionary
$\mathbf{\Psi}\in\Re^{L\times N}$ with $L\leq N$. A signal
$\mathbf{s}$ is said to be sparsely represented under
$\mathbf{\Psi}$ if
\begin{equation}\label{sparse representation}
\mathbf{s=\Psi x},
\end{equation}
Typical examples of the dictionaries include the Fourier matrix for
frequency-sparse signals, a multiband modulated Discrete Prolate
Spheroidal Sequences (DPSSs) dictionary for sampled multiband
signals \cite{zhu2016approximating} and one learned from the
training data \cite{aharon2006rm}. In CS, a sensing matrix
$\mathbf{\Phi}\in\Re^{M\times L}, M\leq L$ \cite{li2013projection,
jiang2017gradient} is utilized to capture most of the information
contained in the signal $\mathbf{s}$ that can be sparsely
represented under $\mathbf{\Psi}$. Thus, for the measurement in
(\ref{cs}) we have $\mathbf{A=\Phi\Psi}$. Algorithms for finding the
sparse vector $\mathbf{x}$ from the measurement $\mathbf{y}$ include
the orthogonal matching pursuit (OMP) \cite{pati1993orthogonal},
basic pursuit (BP) \cite{candes2005decoding}, least absolute
shrinkage and selection operator (LASSO)
\cite{tibshirani1996regression}, iterative hard thresholding
(IHT)\cite{blumensath2009iterative}, dichotomous coordinate descent
(DCD)-based algorithm \cite{zakharov2017low}, adaptive approaches
\cite{saalt,rrecho,rrmser,vfap,sastap,spa,wljio,rdrab,dfjio,doaalrd,locme,okspme,mbdf,wlgmi,damdc,baplnc,kaesprit,tds,sorsvd}
and $\ell_1$-homotopy \cite{donoho2008fast,arh}.

A sparse signal can be successfully recovered when both the values
of the nonzero elements of $\mathbf{x}$ and their positions are
accurately estimated. The use of \textit{a priori} information about
such values and their positions has been shown to provide
improvements to the recovery task \cite{miosso2013compressive} -
\cite{mota2014compressed}. In particular, prior work
\cite{scarlett2013compressed} has exploited the probabilities of the
nonzero elements to improve the recovery using the OMP algorithm.
However, there has been no attempt so far to consider the
improvement of other existing algorithms such as the normalized
iterative hard theresholding (NIHT) \cite{blumensath2010normalized}
with the exploitation of prior knowledge. Therefore, we propose a
knowledge-aided NIHT (KA-NIHT) algorithm and a recursive KA-NIHT
(RKA-NIHT) algorithm, which can exploit prior information in the
form of probabilities of nonzero entries. Note that the
probabilities that have been used in \cite{scarlett2013compressed}
as \textit{a priori} knowledge are fixed. However, in practice, such
a precise knowledge can hardly be available. A more practical
approach is to consider these probabilities just as educated
guesses, and update them within the iterations using the RKA-NIHT
algorithm. To this end, we propose the use of an adaptive set of
probabilities and develop a recursive procedure to compute these
probabilities over the iterations. Simulations show that the
proposed KA-NIHT and RKA-NIHT algorithms outperform existing
recovery techniques.

The organization of the remaining part of the paper is as follows.
Some preliminary work is introduced in Section 2. Section 3
describes the proposed knowledge-aided algorithms using the initial
probability vector and adjustment of the probability vector through
a recursive method based on the NIHT algorithm.  Section 4 presents
and discusses simulation results. Section 5 gives the conclusion of
our work.

\section{Preliminary Work}
\label{sec:Preliminary}

In compressive sensing systems, a designer often employs a recovery algorithm to obtain a sparse solution. A recovery algorithm can be formulated as an optimization problem that can be written as:
\begin{equation}\label{recovery}
\widehat{\mathbf{x}}=\arg\min_{\mathbf{x}:\|\mathbf{x}\|_{0}\leq K}\|\mathbf{y-Ax}\|_{2}^{2},
\end{equation}
where $\|\cdot\|_{2}$ denotes the 2-norm of a vector, Thus, a sparse vector $\widehat{\mathbf{x}}$ should be found so that the error $\|\mathbf{y-Ax}\|_{2}^{2}$ is minimized under the $K$-sparsity constraint $\|\mathbf{x}\|_{0}\leq K$. In the following subsections, the NIHT algorithm and types of \textit{a priori} information on the sparsity are introduced.

\subsection{Normalized Iterative Hard Thresholding}
\label{ssec:IHT}

Under the condition of the restricted isometry property \cite{candes2006robust}, the sparse recovery can be obtained by using the NIHT algorithm \cite{blumensath2010normalized}. In the NIHT algorithm, the cost function decreases in every iteration until the convergence. Let the solution vector be initialized as $\mathbf{x}^{0}=\mathbf{0}$. The following iterations ($l\geq 1$) are used in the NIHT algorithm:
\begin{equation}\label{IHT}
\mathbf{x}_{l+1}=H_{K}(\mathbf{x}_{l}+\mu\mathbf{A}^{T}(\mathbf{y-Ax}_{l})),
\end{equation}
where $H_K(\cdot)$ is an operator setting all elements of a vector to zero except for the K elements with the largest magnitudes. The step size $\mu$ is adapted to maximally minimize the error in every iteration:
\begin{equation}\label{mu}
\mu=\frac{\mathbf{g}_{\Lambda_{l}}^{T}\mathbf{g}_{\Lambda_{l}}}{\mathbf{g}_{\Lambda_{l}}^{T}\mathbf{A}_{\Lambda_{l}}^{T}\mathbf{A}_{\Lambda_{l}}\mathbf{g}_{\Lambda_{l}}},
\end{equation}
where $\Lambda_{l}$ is the support of $\mathbf{x}_{l}$ and $\mathbf{g}=\mathbf{A}^{T}(\mathbf{y-Ax}_{l})$ is the negative gradient of $\|\mathbf{y-Ax}\|_2^2$.

\subsection{\textit{A priori} information on sparsity}
\label{ssec:prior}

The NIHT algorithm exploits \textit{a priori} information o the sparsity of the solution in the form of the number of non-zero elements $K$. The performance of a sparse recovery algorithm can be improved if extra \textit{a priori} information on the sparsity of the solution is available. Let a sparse signal $\mathbf{x}=[x_{1},\cdots, x_{N}]^{T}$ be generated using the following rule. The $i$-th entry of of $\mathbf{x}$ is given by
\begin{equation}\label{x}
x_{i}=w_{i}\vartheta_{i},
\end{equation}
where $w_{i}$ is a deterministic non-zero value, and $\vartheta_{i}\in\{0,1\}$ is a random binary variable used to decide whether the entry is zero or non-zero according to a probability distribution. The probability for $\vartheta_{i}=1$ is denoted as $p_{i}$, and the random variable $\{\vartheta_{i}\}_{i=1}^{N}$ are assumed independent. The support of $\mathbf{x}$ is given by $\Lambda=\{i|\vartheta_{i}=1\}$. In \cite{scarlett2013compressed}, OMP \cite{pati1993orthogonal}, BP \cite{candes2005decoding} and LASSO \cite{tibshirani1996regression} algorithms are extended to exploit \textit{a priori} information in the form of the probability distribution $\mathbf{p}=[p_1, \cdots, p_N]^{T}$ which are named as Log-Weighted OMP (LW-OMP), Log-Weighted BP (LW-BP), Log-Weighted LASSO (LW-LASSO), respectively. The results show that excellent sparse recovery can be achieved if the probability distribution is sufficiently non-uniform, especially using LW-OMP algorithm. Therefore below we will use it as a benchmark for comparison with our algorithm.

The goal of this paper is to propose an efficient algorithm to estimate $\mathbf{x}$ from the observation $\mathbf{y}$ using $\mathbf{p}=[p_1, \cdots, p_N]^{T}$ as prior probabilities based on the NIHT algorithm.

\section{Proposed Reconstruction Algorithms}
\label{sec:new}
The NIHT algorithm according to (\ref{IHT}) selects the support elements with the highest magnitudes. With such strategy, it is possible that non-zero elements with low magnitudes as missed. By exploiting the prior probabilities $\mathbf{p}=[p_1, \cdots, p_N]^{T}$ the support estimate can be made more accurate and consequently the performance of sparse recovery improved.

We propose to recover the support in every iteration of the NIHT algorithm using the following operation:
\begin{equation}\label{support}
\Lambda=\sup[S_{K}(|\mathbf{x}+\mu\mathbf{g}|+\alpha\log(\mathbf{p}))],
\end{equation}
where $S_{K}(\cdot)$ is an operator to sort elements of a vector in descending order, sets last $(N-K)$ elements to zero, and return the elements back to the original order. The operator $\sup[\cdot]$ extracts positions of non-zero elements. Note that the notation $|\mathbf{x}+\mu\mathbf{g}|$ means element-wise magnitudes of the vector elements. Similarly, $\log(\mathbf{p})$ means $\log$ of each element of vector $\mathbf{p}$. The term $\log(\mathbf{p})$ introduces a penalty in the iterations. If $\alpha=0$, then we arrive at the NIHT algorithm. If $\alpha>0$, then for a high $p_i$ (close to one) the penalty is small, while for a small $p_i$ the penalty is high. As a result, if it is \textit{a priori} know that the probability of appearance of the $i$-th element $p_i$ is low, it is unlikely that it would be selected into the support $\Lambda$. The parameter $\alpha$ controls the tradeoff between the importance of the current magnitude $|x_i+\mu g_i|$ of an element and its \textit{a priori} probability $p_i$. The modification of the NIHT algorithm that exploits the support estimate according to (\ref{support}) is named KA-NIHT.

Note that in the KA-NIHT algorithm, the \textit{a priori} probabilities $\mathbf{p}$ are fixed. However, in practice, such a precise knowledge can hardly be available. A more practical approach is to consider these probabilities just as educated guesses, and update them within the iterations using the RKA-NIHT algorithm as described below.

In the RKA-NIHT algorithm, the support estimate (\ref{support}) is replaced with the estimate:
\begin{equation}\label{newsupport}
\Lambda=\sup[S_{K}(|\mathbf{x}+\mu\mathbf{g}|+\alpha\log(\mathbf{q}))],
\end{equation}
when the vector $\mathbf{q}=\mathbf{p}$ at the first iteration and it is updated recursively at the following iterations as:
\begin{equation}\label{recursive}
\mathbf{q}_{\Lambda}\leftarrow\mathbf{q}_{\Lambda}+\beta\mathbf{p}_{\Lambda}.
\end{equation}
Here we denote $\mathbf{q}_{\Lambda}$ a vector which is obtained from the vector $\mathbf{q}$ by only keeping elements on the support $\Lambda$. The parameter $\beta$ is a tuning parameter which is selected according to the experiments.

The steps of the RKA-NIHT algorithm are summarized as follows.

\vspace{0.3cm}
\hrule
\vspace{0.15cm}
Algorithm: RKA-NIHT
\vspace{0.15cm}
\hrule
\vspace{0.1cm}
\small{
{\bf Input}: Data vector $\mathbf{y}\in\Re^{M\times 1}$, compressed matrix $\mathbf{A}\in\Re^{M\times N}$, the sparsity $K$ is known. The probability vector is $\mathbf{p}$. The iteration number is $I$. $c, \kappa, \beta$ are constants.

{\bf Initialization}: the solution vector $\mathbf{x}_{0}=\mathbf{0}$, and
\begin{equation}\label{Alg1}
 \mathbf{x}_{1}=H_{K}(\mathbf{A}^{T}\mathbf{y}), \nonumber
\end{equation}
$\Lambda_{1}=\sup(\mathbf{x}_{1})$, $l=1$; $\mathbf{q}\leftarrow\mathbf{p}$.

{\bf Repeat } until $l>I$:

~~~~~~\emph{Step 1}: $\mathbf{g}_{l}=\mathbf{A}^{\cal T}(\mathbf{y}-\mathbf{Ax}_{l}) $

~~~~~~\emph{Step 2}: $\mu_{l}=\frac{\mathbf{g}^{T}_{\Lambda_{l}}\mathbf{g}_{\Lambda_{l}}}{\mathbf{g}^{T}_{\Lambda_{l}}\mathbf{A}^{T}_{\Lambda_{l}}\mathbf{A}_{\Lambda_{l}}\mathbf{A}_{\Lambda_{l}}}$

~~~~~~\emph{Step 3}: $\Lambda_{l+1}=\sup[S_{K}(|\mathbf{x}_{l}+\mu_{l}\mathbf{g}_{l}|+\alpha\log(\mathbf{q}))]$

~~~~~~\emph{Step 4}: $\hat{\mathbf{x}}_{l+1}=(\mathbf{x}_{l}+\mu_{l}\mathbf{g}_{l})_{\Lambda_{l+1}}$

~~~~~~\emph{Step 5}: \textbf{If} $\Lambda_{l+1}=\Lambda_{l}$ \textbf{then} $\mathbf{x}_{l+1}=\hat{\mathbf{x}}_{l+1}$

~~~~~~~~~~~~~~~~~~~~\textbf{else} $\varrho=(1-c)(\| \hat{\mathbf{x}}_{l+1}- \mathbf{x}_{l}\|^{2}_{2})/(\|\mathbf{A}(\hat{\mathbf{x}}_{l+1}-\mathbf{x}_{l})\|^{2}_{2})$

~~~~~~~~~~~~~~~~~~~~~~~~~\textbf{if} $\mu_{l}\geq\varrho$ \textbf{repeat} until $\mu_{l}\leq\varrho$

~~~~~~~~~~~~~~~~~~~~~~~~~~~~~$\mu_{l}\leftarrow\mu_{l}/(\kappa(1-c))$

~~~~~~~~~~~~~~~~~~~~~~~~~~~~~$\Lambda_{l+1}=\sup[S_{K}(|\mathbf{x}_{l}+\mu_{l}\mathbf{g}_{l}|+\alpha\log(\mathbf{q}))]$

~~~~~~~~~~~~~~~~~~~~~~~~~~~~~$\hat{\mathbf{x}}_{l+1}=(\mathbf{x}_{l}+\mu_{l}\mathbf{g}_{l})_{\Lambda_{l+1}}$

~~~~~~~~~~~~~~~~~~~~~~~~~$\mathbf{x}_{l+1}=\hat{\mathbf{x}}_{l+1}$

~~~~~~\emph{Step 6}: Update $\mathbf{q}$:

~~~~~~~~~~~~~~~~~~$\mathbf{q}_{\Lambda_{l+1}}\leftarrow\mathbf{q}_{\Lambda_{l+1}}+\beta\mathbf{p}_{\Lambda_{l+1}}$

~~~~~~\emph{Step 7}: $l=l+1$

{\bf Output}: $\widehat{\mathbf{x}}=\mathbf{x}_{l}$
}

\vspace{0.15cm}
\hrule

\vspace{0.3cm}

The KA-NIHT algorithm is the particular case of the algorithm RKA-NIHT when step 6 is removed.

The computational complexity of NIHT, KA-NIHT and RKA-NIHT per iteration is calculated and shown in Table \ref{complixity}.

\begin{table*}[htb!]\caption{Computational complexity of three algorithms per iteration.}\label{complixity}.
\vspace{-0.5cm}
\begin{center}
\scriptsize
\begin{tabular}{|c|c|c|c|c|}
\hline
                & Additions     &Multiplications     & Comparison          \\
\hline
NIHT    &$KM+K+2N+3MN+M-5$       &$KM+K+2N+3MN+2M+3$       &$2NlogN+1$         \\

\hline
KA-NIHT    &$KM+K+5N+2.75MN-2.5$       &$KM+K+2.5N+2.75MN+1.75M+0.75$       &$2.75NlogN+1.75K+0.75$         \\

\hline
RKA-NIHT    &$KM+2K+2.6N+2.4MN-1.8$       &$KM+2K+1.8N+2.4MN+1.4M+0.4$       &$2.75NlogN+1.75K+0.75$         \\

\hline
\end{tabular}
\end{center}
\end{table*}

\section{Simulation Results}
\label{sec:simulation}
In this section, we investigate the performance of OMP \cite{pati1993orthogonal}, LW-OMP \cite{scarlett2013compressed}, NIHT \cite{blumensath2010normalized}, KA-NIHT, RKA-NIHT and Oracle NIHT algorithms for various $M$ and noise variance $\sigma^2$ with $N=240$, and $S=1000$ simulation trials.

The values $\{w_i\}_{i=1}^{N}$ in (\ref{x}) are drawn from a Gaussian distribution $N(0,1)$. The support probabilities are modeled as follows. Elements of $\mathbf{x}$ are divided into $G$ groups. The proportions $h_{n}$ of elements in the groups are different, but $\sum_{n=1}^{G}h_n=1$. Every group contains $N_n=h_nN$ elements whose support probabilities are equal to $\overline{p}_n$. For a sparse signal with support $\Lambda$, the sparsity $K$ is the number of non-zero elements $K=|\Lambda|$. The average of $|\Lambda|$ is denoted as $\overline{K}=\sum_{i=1}^{N}p_i=\sum_{n=1}^{G}N_n\overline{p}_n$. In our simulation, the sparse vector is divided into $G=4$ groups with $N_1=210, N_2=20, N_3=5, N_4=5, N_n\overline{p}_n=4$, and $\overline{K}=16$. That is, each group has on average four non-zero coefficients.

We compare the true support $\Lambda_t$ with its estimate $\widehat{\Lambda}_t$ in the $t$-th simulation trail using
\begin{equation}\label{recover}
P_{recovered}=\frac{1}{S}\sum_{t=1}^{S}\frac{|\Lambda_t\cap \widehat{\Lambda}_t |}{|\Lambda_t|},
\end{equation}
Another performance measure adopted in this paper is the Mean Square Deviation (MSD) defined as:
\begin{equation}\label{msd}
MSD=\frac{1}{S}\sum_{t=1}^{S}\frac{\|\mathbf{x}^{(t)}- \widehat{\mathbf{x}}^{(t)}\|_{2}^{2}}{\|\mathbf{x}^{(t)}\|_{2}^{2}},
\end{equation}
where $\mathbf{x}^{(t)}$ is the true vector and $\widehat{\mathbf{x}}^{(t)}$ is the recovered vector in the $t$-th simulation trial.

\begin{figure}[htb!]
\centering
\includegraphics[width=7.5cm]{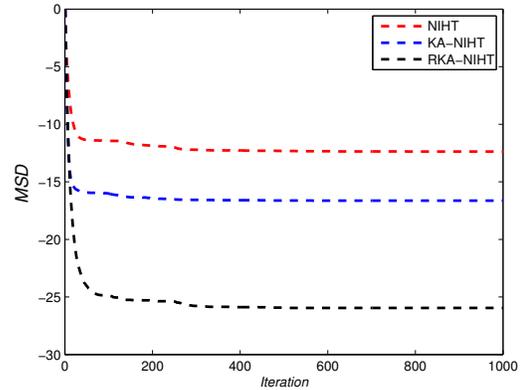}
\vspace{-0.25cm}
\caption{The MSD versus iteration number, with $M=70$;}
\label{iteration}
\end{figure}

Fig. \ref{iteration} draws the MSD evolution with iterations in the three NIHT based algorithms. The variance of noise is $\sigma^{2}=10^{-3}$. The dimension of $\mathbf{y}$ is $M=70$. The other parameters are set to: $c=0.01, \kappa(1-c)=2, \beta=0.6, \alpha=\frac{1.5}{\sum_{i=1}^{N}q_i}$. It can be seen that the proposed algorithms outperform the original NIHT algorithm, with the RKA-NIHT algorithm providing the best performance.

\begin{figure}[htb]

\begin{minipage}[b]{1\linewidth}
  \centering
  \centerline{\includegraphics[width=7.5cm]{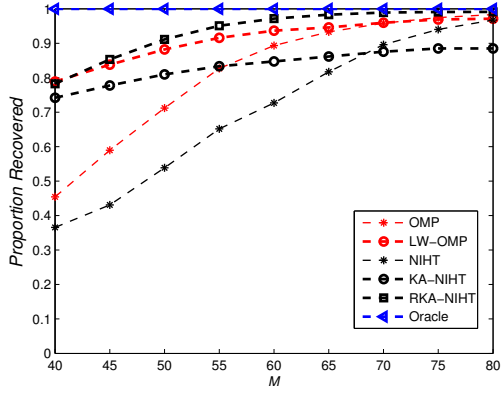}}
  \vspace{0cm}
  \centerline{(a)}\medskip
\end{minipage}

\begin{minipage}[b]{1\linewidth}
  \centering
  \centerline{\includegraphics[width=7.5cm]{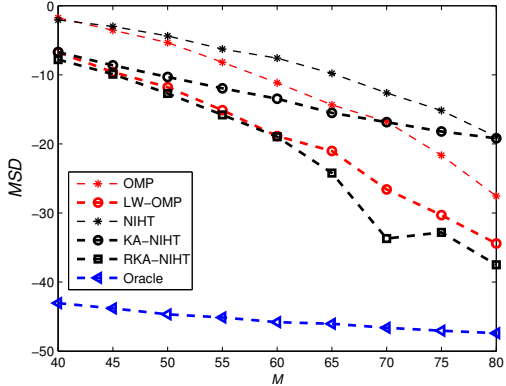}}
  \vspace{0cm}
  \centerline{(b)}\medskip
\end{minipage}

\vspace{-0.3cm}
\caption{The proportion of coefficients recovered and MSD versus sensing matrix
dimension $M$.}
\label{M}

\end{figure}

Fig. \ref{M} (a) demonstrates the proportion of coefficients recovered in six methods with the dimension of the vector $\mathbf{y}$ varying from $M=40$ to $M=80$. Fig. \ref{M} (b) shows the MSD performance.

We can conclude the following:

The proportion of correct recovery increases with the larger dimension of $\mathbf{y}$.

The recovery algorithms with prior information perform better than the original recovery algorithms such as the LW-OMP against OMP and KA-NIHT against NIHT.

As for the KA-NIHT and RKA-NIHT algorithms, the performance of the scheme with adjustable probabilities is better than that with fixed probabilities.

The RKA-NIHT algorithm shows a better performance than the LW-OMP algorithm, also exploiting the \textit{a priori} probabilities.

\begin{figure}[htb]

\begin{minipage}[b]{1.0\linewidth}
  \centering
  \centerline{\includegraphics[width=7.5cm]{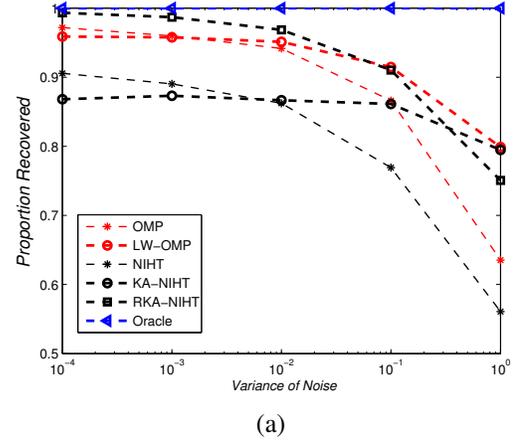}}
  \vspace{0cm}
  \centerline{(a)}\medskip
\end{minipage}

\begin{minipage}[b]{1\linewidth}
  \centering
  \centerline{\includegraphics[width=7.5cm]{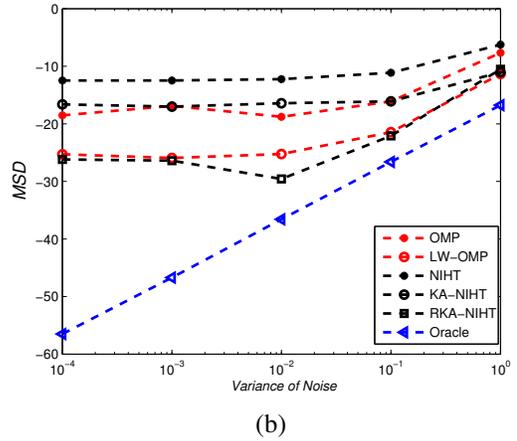}}
  \vspace{0cm}
  \centerline{(b)}\medskip
\end{minipage}

\vspace{-0.3cm}
\caption{The proportion of coefficients recovered and MSD versus variance of noise $\sigma^2$.}
\label{noise}

\end{figure}

Fig. \ref{noise} shows the algorithm performance for varying levels
of noise. The length of $\mathbf{y}$ is $M=70$. Other parameters are
the same as for Fig. \ref{M}. Results in Fig. \ref{noise}
demonstrate that the proposed RKA-NIHT algorithm outperforms the
other algorithms in the large range of the noise level.

\section{Conclusion}
In this paper, we have developed knowledge-aided normalized iterative hard thresholding algorithms for sparse recovery problems. The proposed algorithms, named KA-NIHT and RKA-NIHT algorithms, have been designed considering the initial probabilities which are given at the start as prior knowledge and a recursive procedure to update the probabilities of the non-zero coefficients, respectively. The use of prior probabilities can improve the accuracy of finding the positions for non-zero elements. The simulations have shown that the proposed KA-NIHT and RKA-NIHT algorithms perform very well as compared to existing algorithms.

\vfill\pagebreak

\bibliographystyle{IEEEbib}
\bibliography{ICASSP2018}

\end{document}